\def\fecr{{\rm Fe}$_x${\rm Cr}$_{1-x}$\ }
\def\etal{{\sl et.al.}\ }

\def\PR{ Phys. Rev.\ }
\def\PRL{ Phys. Rev. Lett.\ }
\def\JMMM{ J. Magn. Magn. Mat.\ }
\def\JPCM{ J. Phys. Condens. Mat.\ }
\def\be{\begin{equation} }
\def\ee{\end{equation} }
\def\eq{\ =\ }

\def\n{\noindent }
\def\j{{\bf j}\ }

\documentclass[twocolumn,showpacs,superscriptaddress,floatfix,preprintnumbers,amsmath,amssymb,prb]{revtex4}
\usepackage{graphicx}
\usepackage{dcolumn}
\usepackage{color}
\usepackage{bm}
\begin{document}
\title{Electronic, magnetic and optical properties of random Fe-Cr alloys}
\author{Kartick Tarafder}
\affiliation{S.N. Bose National Center for Basic Sciences, JD-III, Salt Lake City, Kolkata 700098, India }
\author{Atisdipankar Chakrabarti}
\affiliation{R.K. Mission Vivekananda Centenary College, Rahara, Kolkata 700118, India}
\author{Subhradip Ghosh}
\affiliation{Department of Physics, Indian Institute of Technology, Guwahati, India}
\author{Biplab Sanyal}
\affiliation{Theoretical Magnetism Group, Department of Physics,  Uppsala University, SE-75121, Sweden}
\author{Olle Eriksson}
\affiliation{Theoretical Magnetism Group, Department of Physics,  Uppsala University, SE-75121, Sweden}
\author{Abhijit Mookerjee}
\affiliation{S.N. Bose National Center for Basic Sciences, JD-III, Salt Lake City, Kolkata 700098, India }
\date{\today}
\begin{abstract}
In this communication we have studied the electronic structure,
magnetic  and optical properties
 of bcc \fecr alloys in the ferromagnetic phase.
We have used the  augmented space recursion technique coupled
with tight-binding linearized muffin-tin orbital technique (TB-LMTO-ASR) as well as the
coherent-potential approximation based on the Korringa-Kohn-Rostocker
method (KKR-CPA). Also the plane wave projector augmented wave (PAW) method has been used 
with the disorder simulated by   the special quasi-random structure method for
configuration averaging (SQS). This was to provide a comparison between the
different methods in common use for random alloys.  
Moreover, using the self-consistent potential parameters from TB-LMTO-ASR calculations we obtained
 the spin resolved optical conductivity  using the generalized
recursion technique proposed by M\"uller and Vishwanathan.
\end{abstract}
\maketitle

\section{Introduction} 
Multi-layers of magnetic metals have attracted attention because of their
possible applications in designing devices. Fe-Cr multi-layers in particular have
been studied \cite{kub}. It has been felt that in order to understand multi-layers, we must first
attempt to understand binary inter-metallic compounds as well as random binary alloys~:
the former, since in the B2 structure they are naturally occurring models of single atomic
multi-layers,  and the latter, in particular, to give insight into disordering at the multilayer interfaces \cite{qiu}.

There  are ample experimental investigations on this alloy system. These
include studies on structural phase stability and phase diagrams \cite{bren}$^-$\cite{maz}, 
spinodal decomposition   \cite{wil}$^-$\cite{lan}, structural studies 
from X-Ray scattering \cite{bab}$^-$\cite{bab2}, inelastic neutron scattering \cite{kat}$^-$\cite{swan}~,
 M\"ossbauer \cite{chand}$^-$\cite{cie}, heat capacity \cite{down}$^-$\cite{mac},
thermopower \cite{ara},   the magnetic phase stability \cite{MM} and 
 magnetic phases of \fecr alloys  \cite{bur}$^-$\cite{fkk}.
They provide a variety of information, e.g. variation of magnetization with band
filling \cite{al}, moment distribution in dilute Fe based alloys \cite{sw}, 
 composition dependence of high field susceptibility \cite{crsd}, low-temperature 
specific heat \cite{wc} and resistivity anomaly \cite{ns}. 

Theoretical investigations have been carried out on the phase stability of \fecr alloys 
\cite{rao}$^-$\cite{batav}. There have been several calculations on ordered inter-metallic Fe-Cr in the
B2 structure using standard electronic structure methods \cite{qiu}$^,$\cite{mmp}$^-$\cite{singh}.
Studies on Fe-Cr in the $\sigma$ phase have also been reported \cite{sluiter}. 

Magnetism in this alloy has been studied by the spin polarized KKR-CPA method
 by many authors \cite{moroni}$^,$\cite{moriatis}. Butler \etal \cite{but} have studied
the GMR effect
 in the concentration range of (0.5$<x<$0.1) and antiferro- to ferro-magnetic transition
at the critical concentration of $x=0.3$.
Dederichs \etal \cite{ded} have discussed the Slater-Pauling curves of \fecr.
Kulikov \etal \cite{kul} have shown that body-centered cubic Fe moments are
fairly independent of Cr concentration, while the opposite is
true for Cr.

Jiang \etal \cite{jiang}  have studied the local environment
effect on the formation enthalpy, magnetic moments, equilibrium
lattice parameter and bond lengths using special quasi-random structure (SQS), a concept
proposed by Zunger \etal \cite{zunger}. With a 16-atom SQS super-cell they have shown
that even for a lattice-matched system like Fe-Cr the average
Cr-Cr, Cr-Fe, Fe-Fe bond lengths are quite different.
For magnetic moment calculation they have obtained reliable results
in the concentration range of $x > 0.3 $.
Very recently, Olsson \etal \cite{olsson} have studied the stability of FeCr alloys in the ferromagnetic phase. 
They argued that the negative mixing enthalpy responsible for this stability has an electronic origin. They also showed that
 PAW-SQS calculations of mixing enthalpy and magnetic moments were in very good agreement with CPA coupled with the exact-muffin-tin-orbi
tal method.

It is evident from our introductory remarks and the wealth of references given, that the \fecr alloy
system has been studied thoroughly both experimentally and theoretically for quite some time now.
There can be no justification for one more calculation using one more method at this stage.
However, what has not been done so far is a systematic analysis and comparison of the results
of the principal successful methodologies. Such a study will  give us a clear picture of the comparative
strengths and weaknesses of these methods vis-a-vis one another. An extensively experimentally
studied alloy, like  \fecr, is then an excellent choice of a system on which to carry out such a 
comparative analysis. This is the basic aim of the work presented here. 

We shall identify 
a few of the first-principles electronic structure methods for disordered alloys which we
believe to be the most accurate and make a comparative analysis of results obtained through
them for \fecr. Different properties like optical response and magnetic transition temperatures
have rarely been studied using the same electronic structure methods coupled with different
approximations dealing with disorder. 
This will provide insights into the advantages and drawbacks of these  techniques and give us  
confidence in their use for future studies on different alloy systems.

We have identified  three   electronic structure methods for disordered substitutional binary alloys : the Korringa-Kohn-
Rostocker based mean field coherent potential approximation (KKR-CPA) \cite{kkrcpa}, the projector augmented wave based
super-cell calculations on special quasi-random structures (PAW-SQS)\cite{zunger} and    the tight-binding
linear muffin-tin orbitals based augmented space recursion (TB-LMTO-ASR) \cite{tf}.  The KKR and its linear version,
LMTO, are among the accurate techniques in use for the study of random substitutional alloys, where
the disorder induced scattering is {\sl local}.
 Of course, LMTO being a linearized
approximation to the KKR, is expected to be less accurate. However, if the energy window around
the linearization energy nodes is not too large, the LMTO estimates energies with tolerances of
around 5-10 mRy. Unless we are interested in estimating energy differences smaller than
this quantity or over large energy windows, the LMTO is good enough and has the great advantage that its secular equation is
an eigenvalue problem  rather than the more complicated functional equation of the KKR. 
In PAW calculations, we have the advantage of a full potential without any shape approximation.
 It has already been shown that electronic structures of alloys are quite well described with
 the SQS method \cite{sqsall}.
 We shall use all three of these methods and compare and contrast the corresponding results.

Methods to deal with disorder fall into three categories. In the first category
belongs  the single-site mean-field
CPA. This approximation has been eminently successful in dealing with a variety of disordered
systems. However, whenever there is either strong disorder fluctuation scattering, as in dilute,
split-band alloys or when local environment effects like short-ranged ordering,  clustering and segregation,  or 
local lattice distortions due to size mismatch of the constituent atoms become important, the single-site based CPA becomes inadequate. 

In the
second category belong the generalizations of the CPA, of which, the augmented space based
methods : the itinerant CPA \cite{icpa} (ICPA) and the augmented space recursion (ASR) \cite{asr}, are foremost. They not only
retain the necessary analytic (Herglotz\cite{her}) properties of the averaged Green function, as the CPA does, but
also properly incorporate local environment effects. 
 
In the third category belong the super-cell based calculations. Zunger \cite{zunger} suggested that if we construct
a super-cell and populate its lattice points randomly by the constituents so as to mimic the
concentration correlations in the random alloy, a single calculation with this super-lattice should
approximate the configuration average in the infinite random system. The SQS 
 approach has been used to incorporate short-ranged order and local lattice distortions
in alloy systems. Certainly, in the limit of a very large super-cell this statement is the
{\sl theorem of spatial ergodicity}. This theorem provides  the explanation of why a single experiment on global
property of a bulk material  most often produces the configuration averaged result, provided the
property we are looking at is {\sl self-averaging}. How far this
approach is accurate with a small cluster of, say, 16 atoms, is a priori uncertain. We shall
use the SQS method for averaging as well and compare this with our mean-field and ASR results.

 Before we proceed further, we should note that all the three methods, described earlier, to deal with disorder
are essentially real space approaches. The disorder is substitutional and {\sl at a site}. There is a fourth
technique recently introduced based on a reciprocal-space renormalization method : the non-local CPA.
 The basic technique was
introduced by Jarrel and Krishnamurthy \cite{jarr} and applied to CuZn alloys by Rowlands \etal \cite{row}.
The method is capable of taking into account environmental effects and short-ranged ordering. As expertise
in this method was not available with us we could not include it in this work. Once work on FeCr is available
using this technique, it would be interesting to compare the results with what we project here.

 A systematic comparison between these calculations
 will allow us to ascertain, for example,  whether the CPA is indeed inadequate in some situations
 and how far  the augmented space based techniques improve matters.
We shall extend the TB-LMTO-ASR method also to study the optical response of \fecr alloys. We
are not aware of experimental studies of optical conductivity on this  alloy system and our
preliminary results will provide incentive for further experimental studies.

Ghosh  \etal \cite{ghosh}  have earlier studied the same alloy system.
Their approach was different from  our ASR work here. In the earlier work the authors  used the charge neutral sphere approximation proposed by Kudrnovsk\'y \etal \cite{kuddu} to bypass any
 Madelung contribution to the total energy. This procedure is both
cumbersome and, as we discussed in an earlier paper \cite{durga}, such volume conserving
charge neutral spheres may not necessarily be found. Our present work uses
the ideas of Ruban and Skriver \cite{rusk} to estimate the Madelung energy of charged atomic
spheres during the LDA self-consistency loops.
 This earlier work did not  observe any sign of the experimentally
observed reversal of the Cr projected magnetic moment as a function of Fe concentration.

\section{Electronic  and magnetic structure of  \fecr alloys}

\subsection{Density of States}

\begin{figure*}[p]
\centering
{\resizebox{5in}{4.5in}{\includegraphics{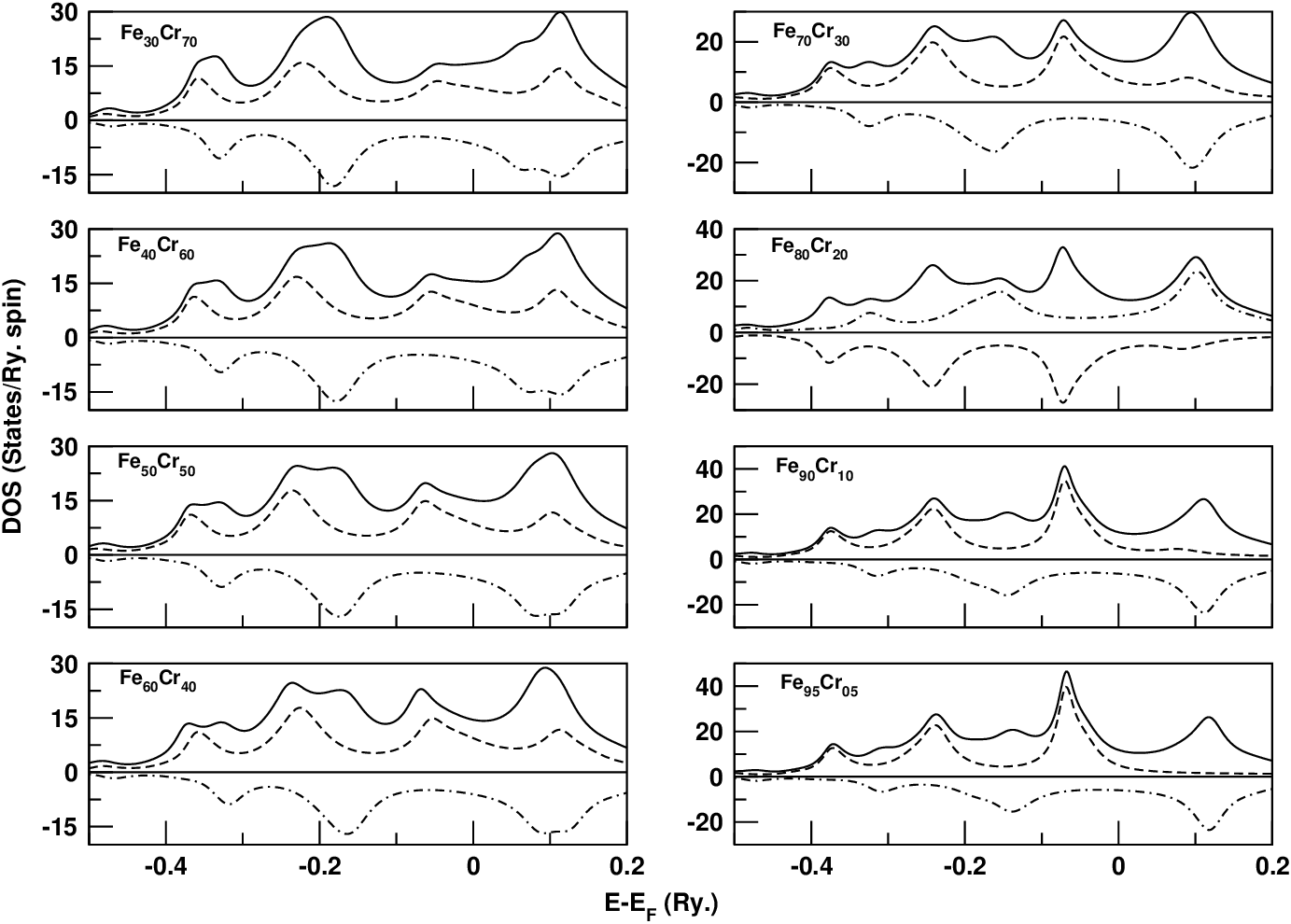}}}
\vskip 2cm
{\resizebox{5in}{4.5in}{\includegraphics{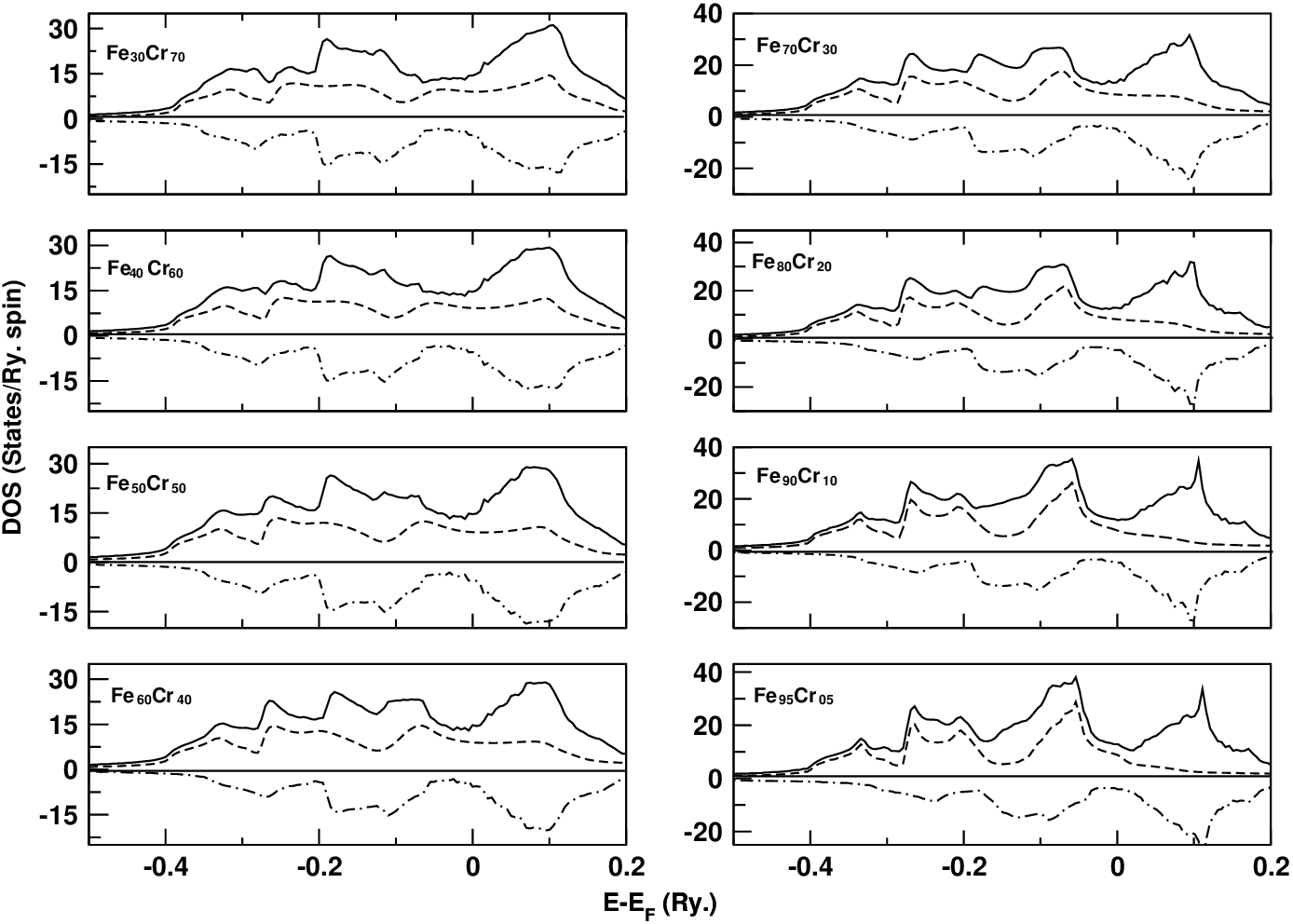}}}
\caption{\label{dos} Densities of states for the spin-up states (upright curves in dashed lines)
spin-down states (inverted curves with dashed lines) and total density of states (full lines) for \fecr 
for various compositions in the ferromagnetic, bcc, disordered phase. The energies
are shown with respect to the Fermi level as the reference level. (top) TB-LMTO-ASR 
(bottom) KKR-CPA. }
\end{figure*}

\begin{figure*}
\centering
\resizebox{4.5in}{2.0in}{\includegraphics{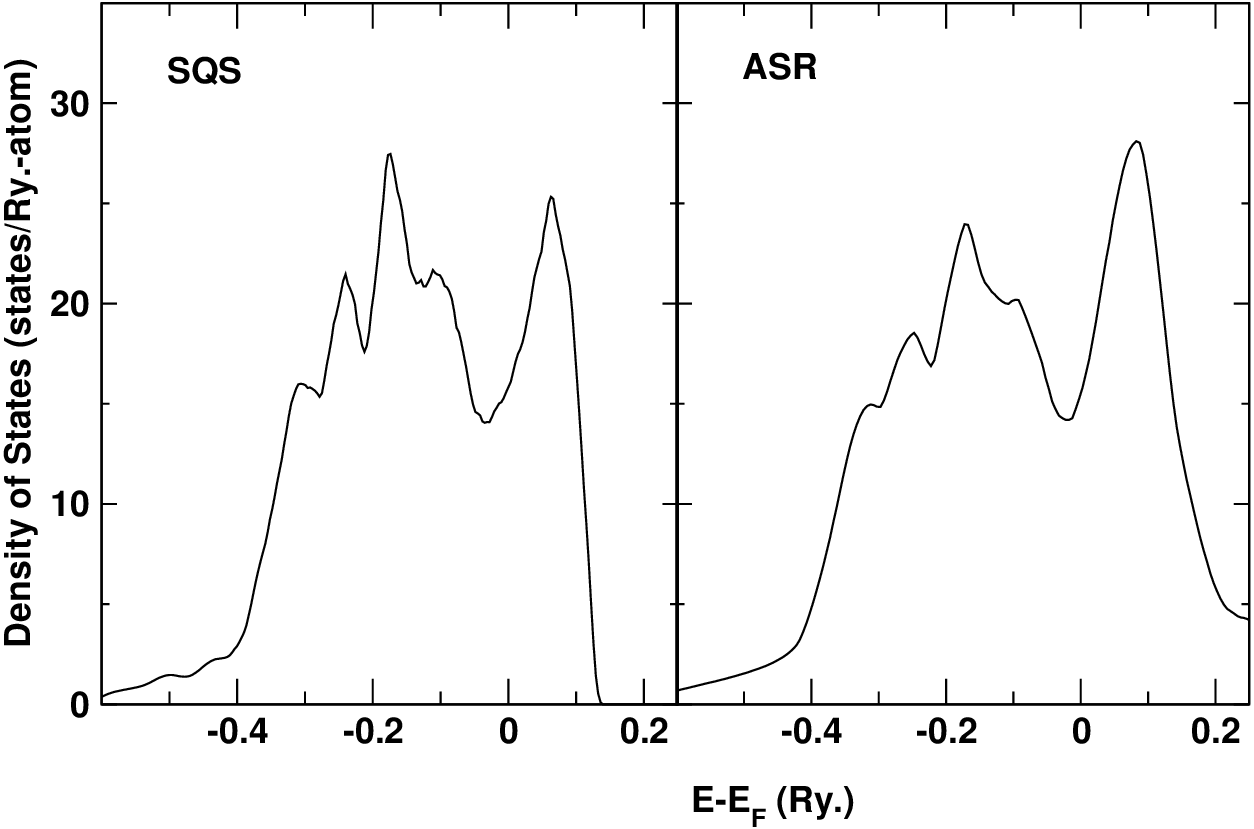}}
\vskip 3cm
\resizebox{4.5 in}{5in}{\includegraphics{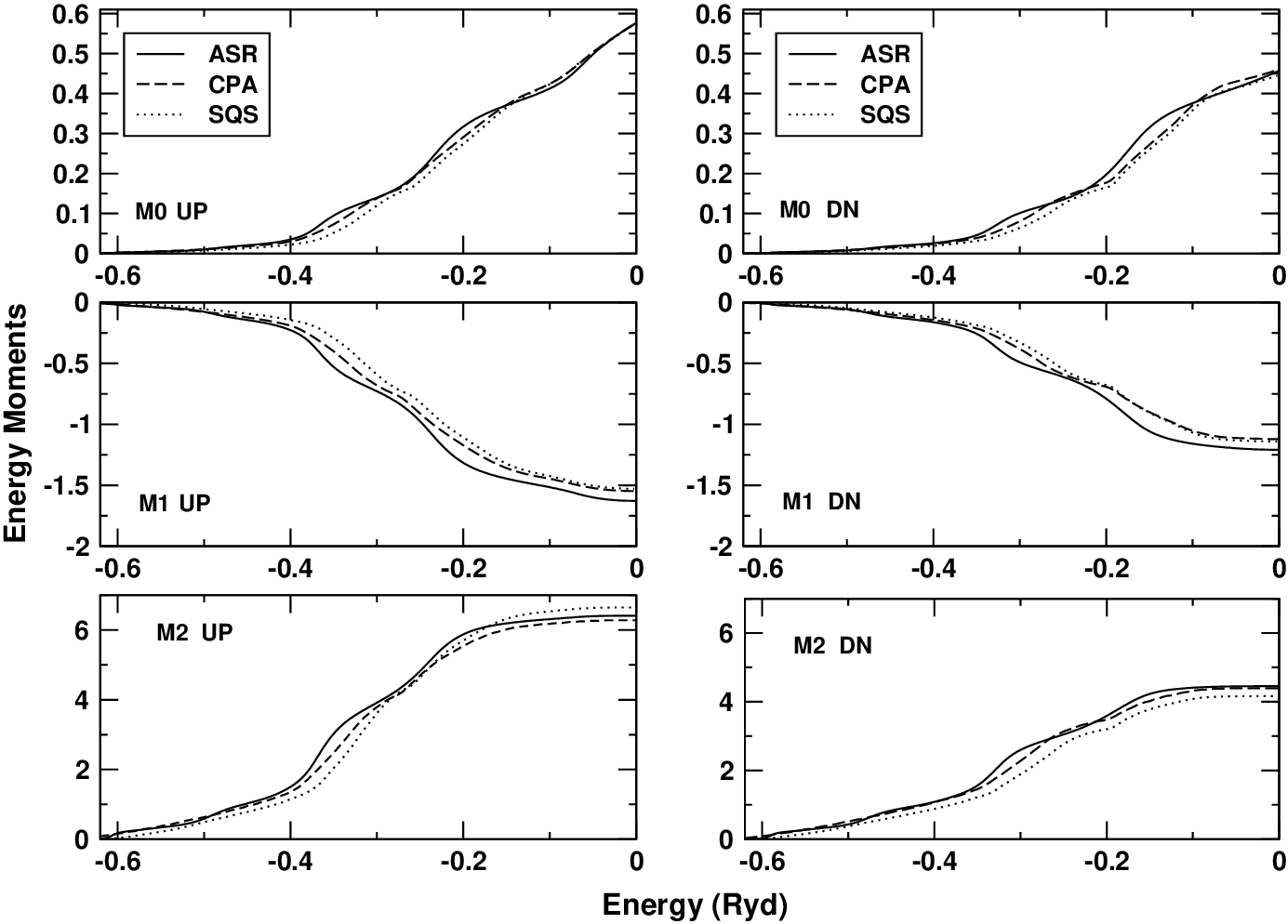}}
 \caption{\label{asrsqs} (top) Densities of states for Fe$_{50}$Cr$_{50}$ obtained 
from the TB-LMTO-ASR and PAW-SQS methods.
(bottom) The first three energy Moments of the density of states (M0, M1 and M2) are shown for the 
TB-LMTO-ASR, KKR-CPA and PAW-SQS, for up spin (UP) and down spin (DN).}
\end{figure*}

Fig. \ref{dos} compares the density of states for the spin-up and spin-down states for \fecr 
across the composition range $0.3\leq x\leq 0.95$ in which the solid solution in the bcc lattice and
the ferromagnetic state is stable at low temperatures. Both the KKR-CPA and TB-LMTO-ASR were
carried out within the LSDA self-consistency with the same Ceperley-Alder exchange correlation functional with  Perdew-Zunger parametrization. The densities of states for both the approaches
show remarkable similarities except in the low Cr compositions. 
 This is expected. The $d$-band
centers of Fe and Cr are well separated, in such a situation, it has been known that for
the dilute concentrations the CPA is not accurate. Originally, it was to reproduce this
regime of parameters, i.e. dilute limit in the {\sl split band} case that the generalizations to the CPA were proposed. The effect here is, however, small as compared with our earlier analysis of Cu$_x$Zn$_{1-x}$\cite{cuzn}.

\begin{figure}
\centering
\includegraphics[width=3in,height=2.0in]{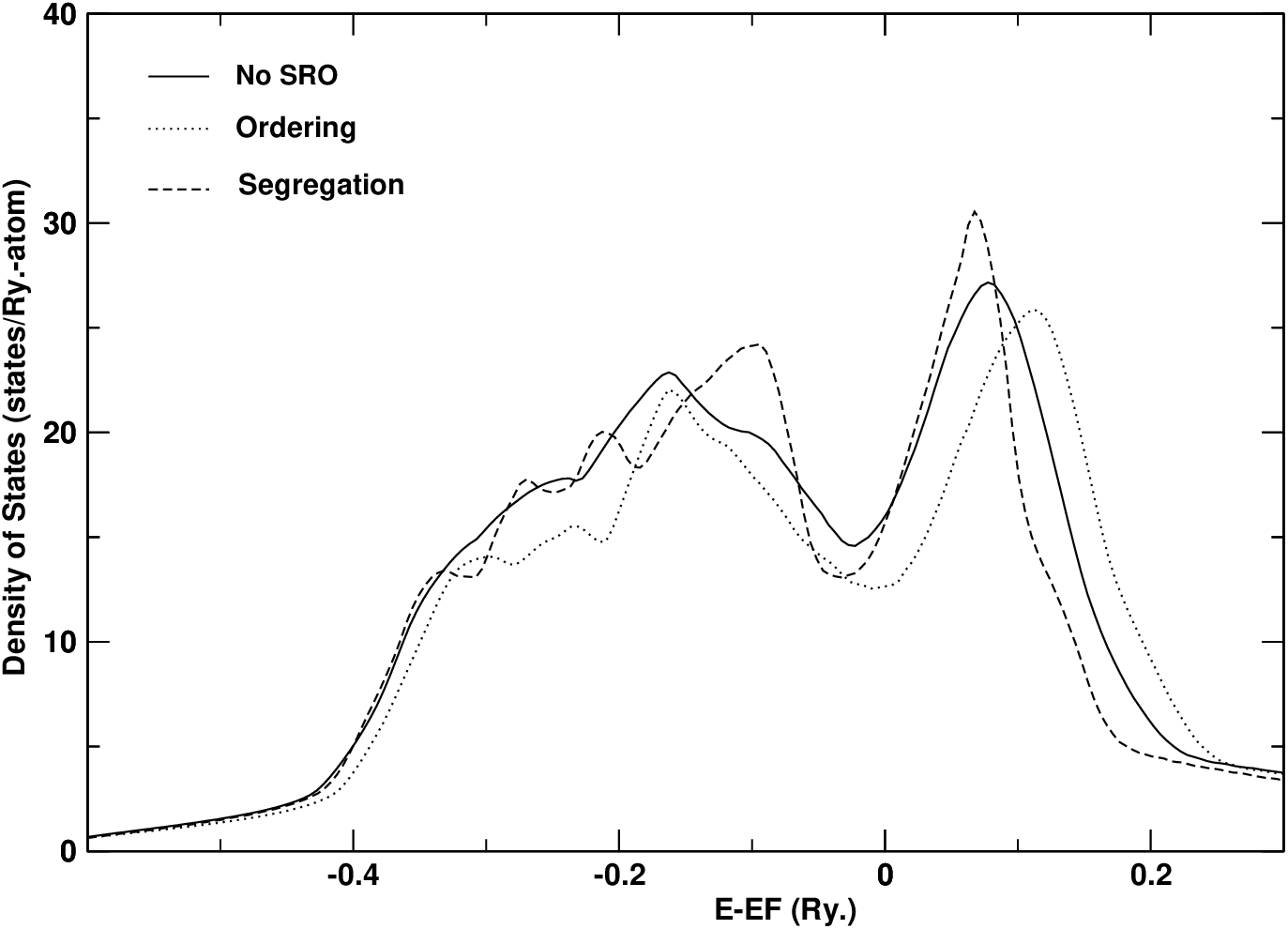}
\vskip 0.8cm
\includegraphics[width=3.in,height=2.0in]{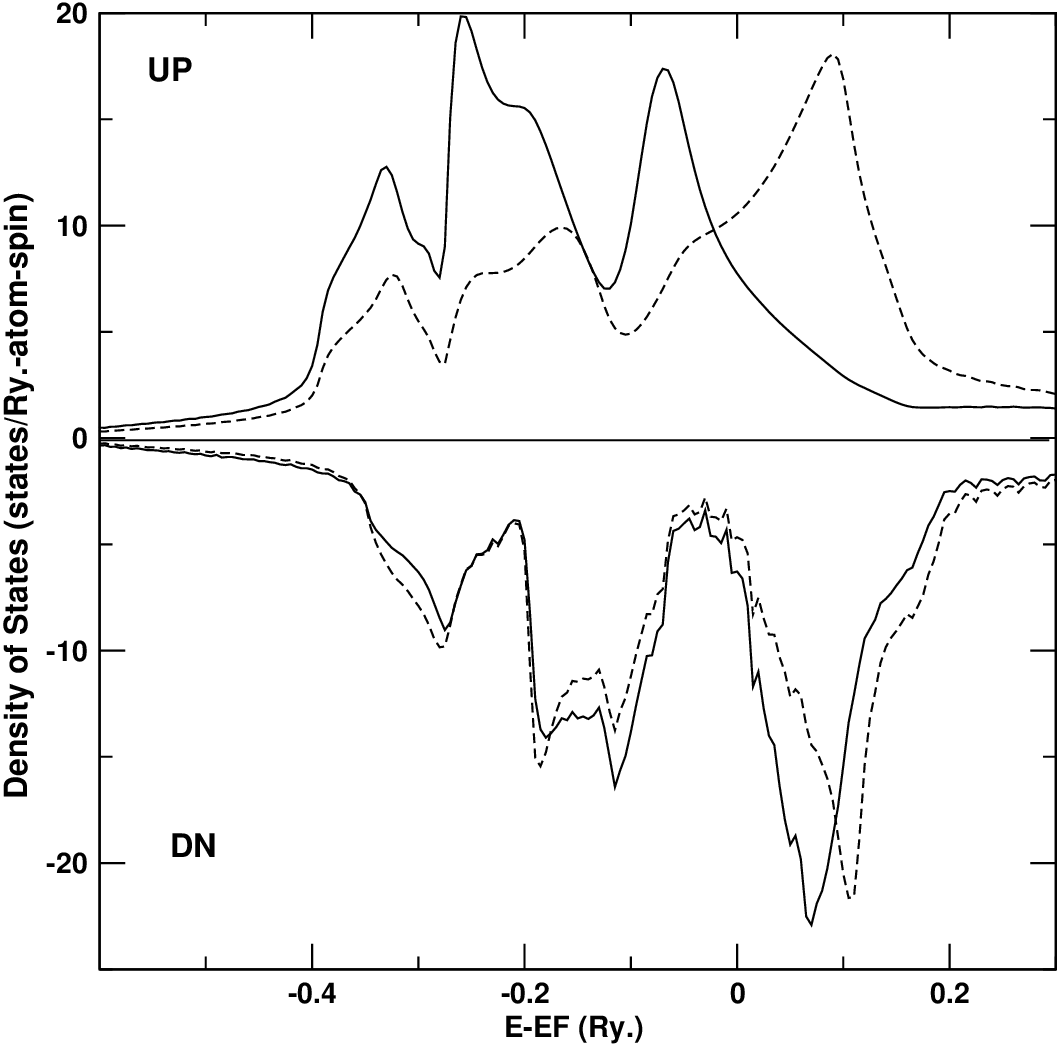}
\caption{(top) Density of States for Fe$_{50}$Cr$_{50}$ for (full line)  $\alpha$=0 (dotted line) $\alpha$=-1
(dashed line) $\alpha$=1. (bottom) Component and spin projected density of states for Fe$_{50}$Cr$_{50}$ for $\alpha$=0. Fe is shown in full lines and Cr in dashed lines. Upright curves are for the up state and inverted curves for the down states.}
\label{sro1}
\end{figure}

Of course there are differences in detail. The KKR-CPA is a reciprocal space based
approach while the TB-LMTO-ASR is a real space based one and the approximations in the two cases are
different. The KKR-CPA is based on a single-site mean field and relevant 
Brilluoin zone integration involves techniques like the tetrahedron integration. The TB-LMTO-ASR  expands the configuration averaged Green function as a continued fraction and the main approximation involves calculation of its asymptotic
``terminator". As compared with an earlier work on Cu$_x$Zn$_{1-x}$ \cite{cuzn}, where the ASR showed considerable
improvement over the CPA, in this particular alloy system the differences are much less prominent, except in the
very dilute limit. This is an important observation, and we should be careful in making general and strong statements about
the efficiency of one method over the other. The results are strongly system dependent.

\begin{figure}
\centering
\vskip 0.8cm
\includegraphics[width=2.5in,height=3.in]{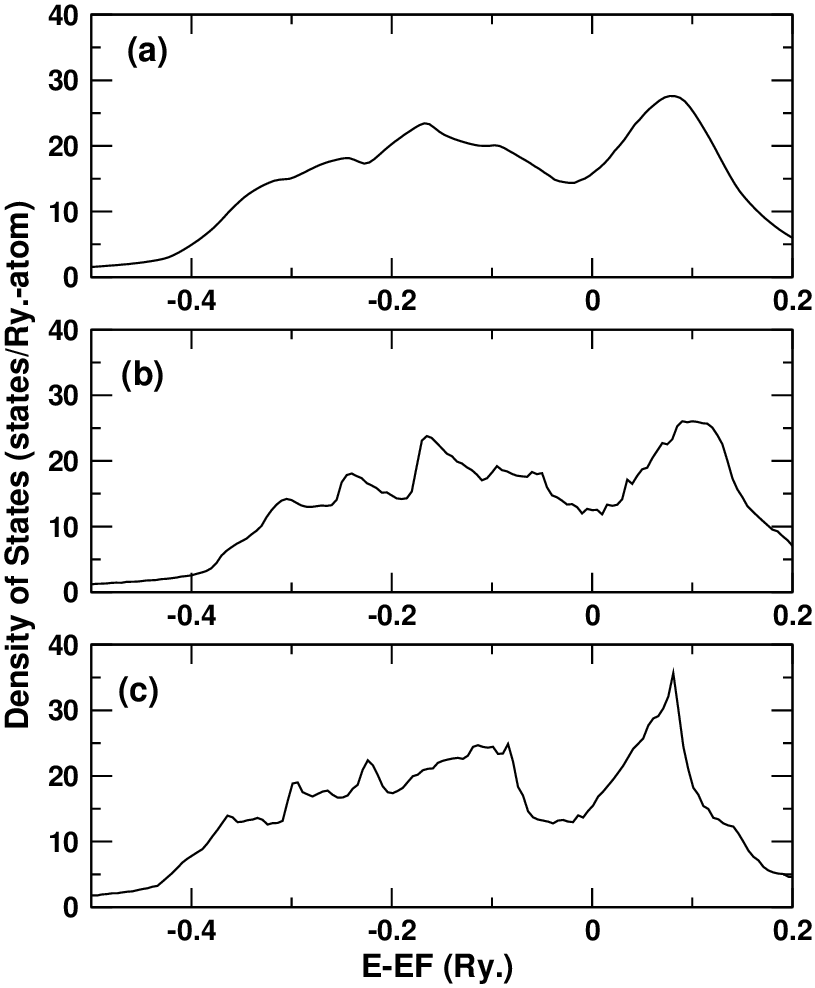}
\caption{(top) Density of States for disordered Fe$_{50}$Cr$_{50}$,  
(middle) Density of States for L12 ordered Fe$_{50}$Cr$_{50}$,
(bottom) Density of States for pure Fe and Cr summed up.}
\label{sro2}
\end{figure}

For the PW-SQS, we have compared the density at only the 50-50 concentration with TB-LMTO-ASR. This is shown in Fig. \ref{asrsqs}.
 This
calculation has been done with the plane wave PAW method implemented in the VASP code \cite{vasp}. 
400 eV was considered as the cut-off energy for the basis set.
 The SQS does simulate disorder, nevertheless, being a super-cell technique,
 it satisfies Bloch's Theorem. 
Therefore, the complex self-energy which arises due to disorder scattering in the ASR is per se absent in SQS calculations. 
This ``life-time" effect smoothens the structures in the ASR density of states and gives them    larger
widths as compared to the SQS. Therefore, to compare the SQS density of states with the TB-LMTO-ASR or KKR-CPA we have to convolute it with a small Gaussian broadening.

Haydock \cite{hay_35}, in his critique of the recursion method, argued that the density of states is perhaps not the
best property to compare between different approximations, because it is unstable to small 
 perturbations. Thouless \cite{thou} has argued that the spectral density arising out of extended states
in a disordered system is extremely sensitive to small perturbations. 
Minor differences in approximations lead to relatively large changes in the spectral density. Haydock suggested that 
it  would be more proper to compare integrated functions like $M(E)\ =\ \int^E_{-\infty} dE' f(E') n(E')$ where $f(E')$ is a well-behaved function of $E'$. Examples are the integrated density of states,
the Fermi  and  band energies and the various energy moments of the density of states. 
In fact, most physical properties are integrated functions of this kind. 

The Fig. \ref{asrsqs} (bottom)  shows the first three moments of the density of states. The moments
have been calculated with normalized densities of states, so that the integrated density
of states (which is the zero-th energy moment of the density of states) is 1 at E=E$_F$
for all the three approximations. The moments match well throughout the energy range up to
the Fermi energy with the relative deviations being no more than 10$\%$ throughout the energy
range of interest. Simply from the density of states there is little to choose between the
different methods. 
\begin{figure*}
\centering
\vskip 1.4cm
\resizebox{1.8in}{2.in}{\includegraphics{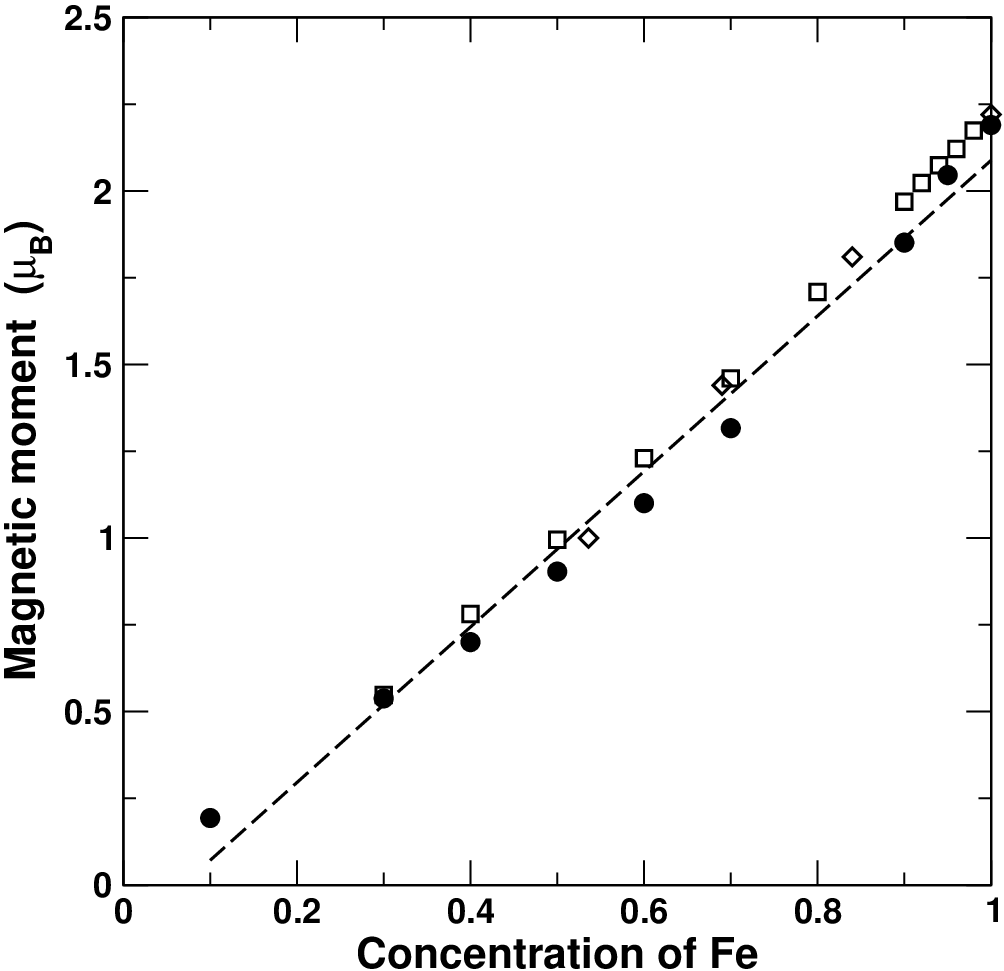}}
\hskip 0.5cm
\resizebox{1.8in}{2.in}{\includegraphics{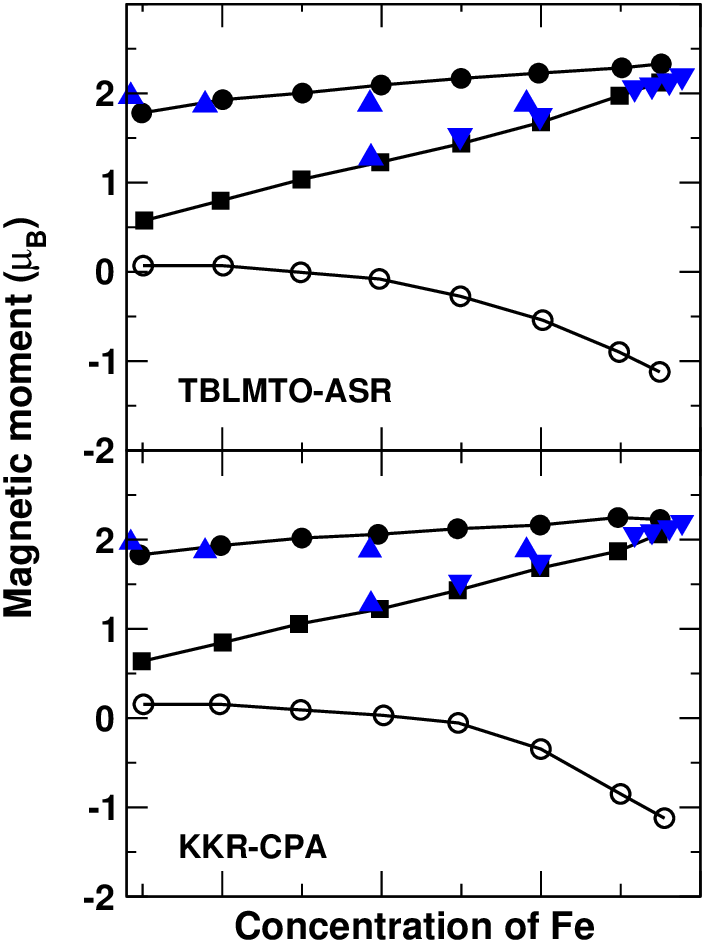}}
\hskip 0.5cm
\resizebox{1.8in}{2.in}{\includegraphics{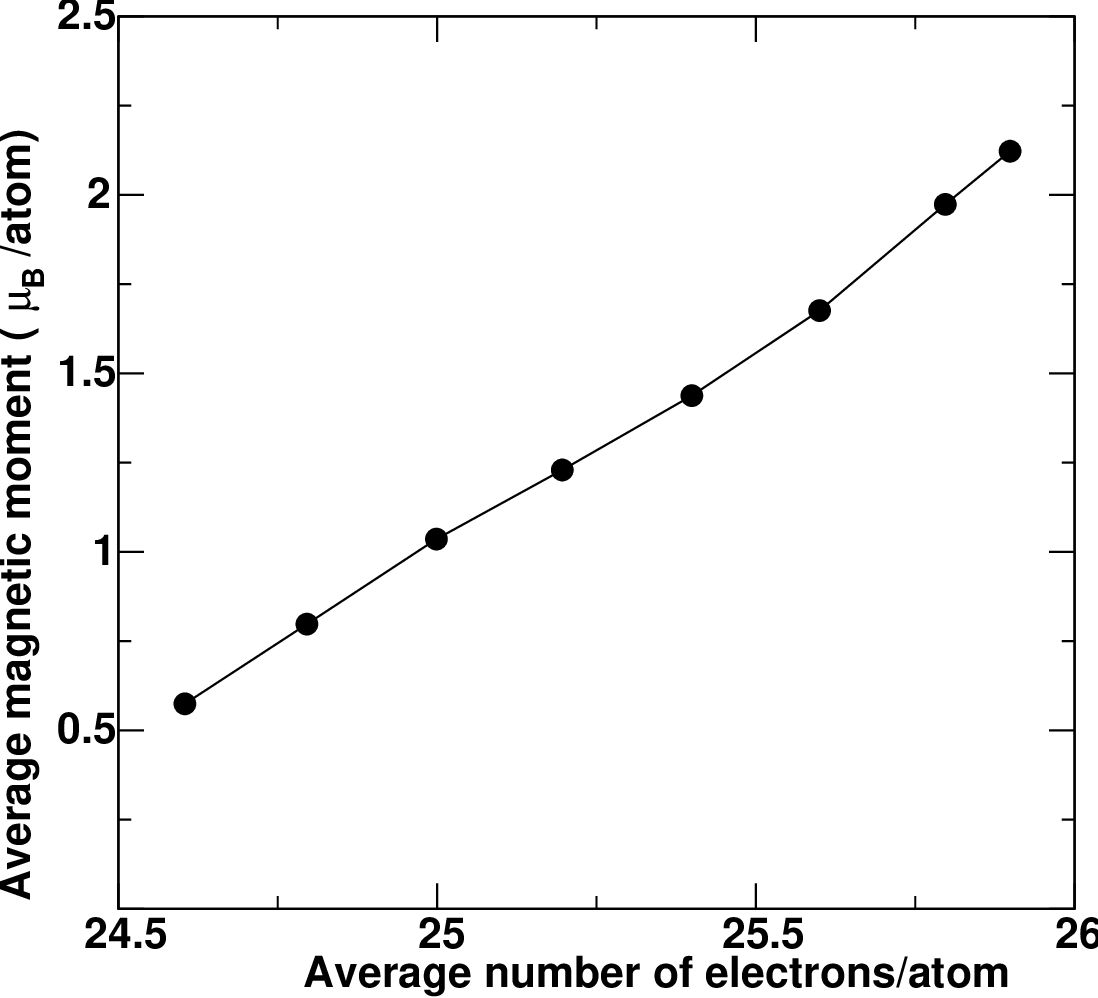}}
\caption{\label{magmom} (left) Compendium of experimental data on 
the averaged magnetic moment as a function of composition. (middle) Local magnetic moments on Fe (filled circles) and Cr (open circles) and
the the average magnetic moment in the alloy (closed squares) for different 
concentrations of Fe using TB-LMTO-ASR (top panel)
and  using KKR-CPA (bottom panel). The available experimental data are shown
as triangles (bottom) Variation of average magnetic moment with 
the average number of electrons using TB-LMTO-ASR.}
\end{figure*}
\begin{figure*}
\centering
\vskip 1.4cm
\resizebox{5in}{5in}{\includegraphics{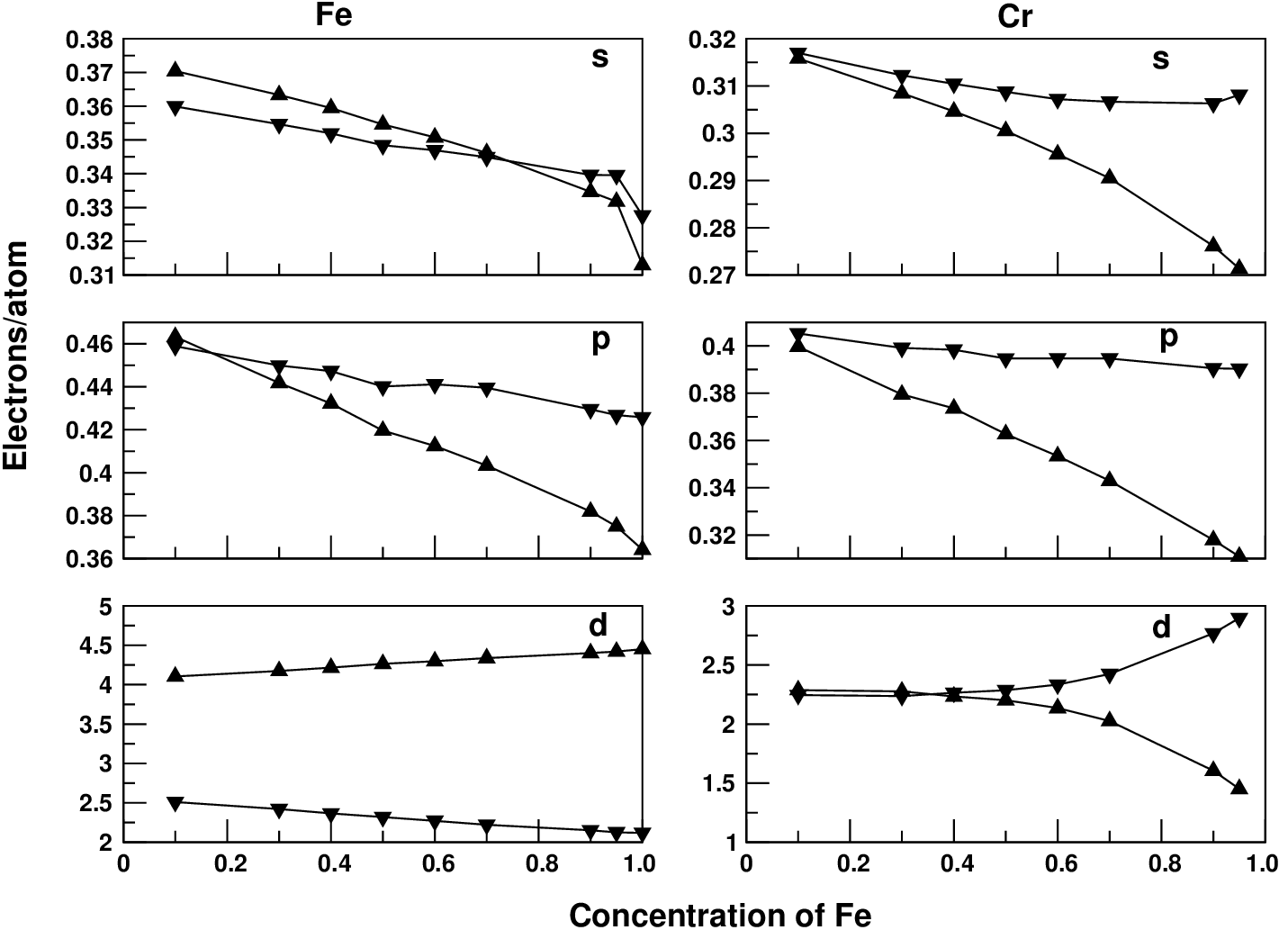}}
\caption{\label{charge} The orbital resolved changes in the Fe and Cr atomic spheres in \fecr
alloys}
\end{figure*}
\subsection{Short-Ranged Ordering.}

The phase diagram of Fe-Cr is simple at high temperatures \cite{elliot}
and a complete range of bcc solid solutions exists from 1093K to the solidus. Alloys
quenched from these temperatures retain their bcc structure. However, the alloys are
not homogeneously disordered. Neutron scattering experiments \cite{al} indicate
that short-ranged clustering exists in these quenched alloys leading to a miscibility
gap at temperatures lower than 793K. The single site CPA cannot deal with short-ranged
order as the latter explicitly involves independent scattering from more than one site. 
Attempts to develop generalizations of the coherent potential approximation (CPA) including effects
of short-ranged order (SRO) have been many,  spread over the last several decades. Many of them
fail the analyticity test.  Mookerjee and Prasad \cite{mp}
generalized the augmented space theorem to include correlated disorder. However, since they then went on to apply it in the cluster CPA approximation, they could not go beyond the two-site cluster and that too only in model Hamiltonians.
The breakthrough came with the augmented space recursion (ASR) approach proposed by Saha \etal \cite{sdm1}$^-$\cite{sdm2}. The method was a departure from the mean-field approaches which always began by embedding a cluster
in an effective medium which was then obtained self-consistently. As discussed earlier, here the Green function was expanded in a continued fraction whose asymptotic part was obtained from its initial steps through an ingenious {\sl termination}
procedure \cite{hay_35}. In this method the effect at a site of quite a large environment around it could be taken into account depending how far one went down the continued fraction before {\sl termination}.
The technique was made fully LDA-self-consistent within the TB-LMTO 
approach \cite{asr} and several applications have been carried out to include short-ranged order in different alloy systems \cite{durga}. 
Details of the formalism has been described in detail in an earlier paper \cite{cuzn}. Here we 
shall apply it to the case of 50-50 FeCr alloys.

We have carried out the TB-LMTO-ASR  calculations on FeCr including short-ranged ordering described by
 the nearest-neighbour Warren-Cowley parameter $\alpha\  (-1\leq\alpha\leq 1)$. The Fe and Cr potentials are self-consistently
obtained via the LSDA self-consistency loop. All reciprocal space integrals
are done by using the generalized tetrahedron integration for disordered systems introduced by us earlier \cite{kspace}. 

 To discuss the effect of SRO, leading, on one hand, to ordering ($\alpha < $ 0)
 and  segregation on the other ($\alpha >$ 0), let us look at Figs. (\ref{sro1}) -(\ref{sro2}).
  The component projected density of states with the completely disordered alloy (Fig. \ref{sro1}, bottom)  shows
rather interestingly that for the down-spin electrons the positions of the centers of the $d$-bands of
Fe and Cr are almost degenerate and strongly hybridize. However, the $d$-bands of the up-spin
electrons are  separated in energy. FeCr is then a partially split-band alloy.
 This implies that for the up-spin electrons the ``electrons travel
more easily between Fe or between Cr sites than between unlike ones" \cite{row}. So when the alloy orders and unlike sites sit next to each other, the overlap
integral between the unlike sites is lower, and for up-spin density of states narrow. For the down-spin
bands this effect is not present.

Fig (\ref{sro1}) (dashed curve) shows the density of states with  $\alpha = 1$. Positive $\alpha$ indicates a clustering or segregating tendency. Comparing with Fig. (\ref{sro2},c), which is a direct sum of
the density of states of Fe and Cr, we note that, with clustering, the density of states begins
to show the structures seen in the pure metals. For $\alpha = 1$ there is still residual long-ranged disorder. This causes smoothening of the bands with respect to the pure materials. For segregation we notice
a shift in the Fe and Cr based structures in the density of states. Again, there is greater smoothening
of the density of states structures when the disorder is perfect Fig (\ref{sro1}) (full lines). This is
due to enhanced disorder scattering leading to larger self-energies.

Fig. (\ref{sro1}) (dotted curve) shows the density of states with $\alpha = -1$ which indicates 
 nearest neighbour ordering. On the fcc lattice at 50-50 composition we expect this ordering 
to favor a L12 structure. We can compare with the density of states for the L12 structure
shown in Fig. (\ref{sro2}). There is a upward shift of the Cr based feature and, as expected, there is
less smoothening than the completely disordered case. We have to realize that as we have taken only
the nearest neighbour short-ranged order, $\alpha = -1$ does not imply perfect long-ranged ordering.
 
\subsection{Magnetic moments}

	Experimental work on  magnetism in \fecr alloys has a long history.
The very earliest works with  \fecr are that of Fallot \cite{fal} in 1936 on the variation of the  averaged magnetic moment with composition and Shull and Wilkinson \cite{sw} on the neutron-scattering  study of Fe-rich \fecr in 1955. Among other properties these studies gave estimates of both the averaged magnetic moments and atom-projected local magnetic moments in this alloy system. Matthews and Morton \cite{MM} studied FeCr and suggested co-existence of ferro and antiferro-magnetism in the alloy, an idea which was not taken up subsequently. Bulk
magnetization measurements were carried out by Aldred \cite{al} in 1976. The
author also examined theoretical explanations for the averaged magnetic moments in the alloy system and concluded that empirical tight-binding CPA estimates of
Hasegawa and Kannamori \cite{has} and Frolliani \etal \cite{fro} gave adequate
description of his experimental data. A series of theoretical approaches using
different electronic structure methods and usually the CPA followed : notable
among these were the works of Moroni and Jarlborg \cite{moroni}, Singh \cite{singh}, Moriatis \etal \cite{moriatis} and Qiu \cite{qiu} whose ``fixed moment" method
yielded a rather large local moment on Cr compared to the general consensus.
Finally, Cie\'slak \etal \cite{crsd} gathered together results for both the
Curie temperature and the averaged magnetic moment from resistivity minima 
\cite{ns}, specific heat anomaly \cite{wc} and elastic measurements \cite{demoron}. A compendium of the
 acceptable experimental results on the averaged magnetic moments for different compositions is shown in Fig. (\ref{magmom}, top left). Results from different experiments are shown by different symbols. The TB-LMTO-ASR  theoretical
results which fit the curve $m(x)\ =\ 2.44\ x\ -\ 0.244$ is shown by the dashed curve. The KKR-CPA prediction for the
 total magnetic moment per atom is almost the same. The theoretical predictions from both the methods agree
very well with the experimental results.

 Figure \ref{magmom} (top middle and right) shows the variation of the local
and average magnetic moment as a function of Fe concentration of the
alloy studied. The results are in good agreement with the few experimental
observations available. It shows that the local Fe moment with
increasing Fe concentration remains almost constant over the
entire concentration range while the Cr moment changes its sign
($x > 0.4$) from low positive value to very high negative value. Earlier
studies have also observed similar behavior. The KKR-CPA and TB-LMTO-ASR
are in good agreement for the Fe local magnetic moment, while the agreement is
less close for the local moment on Cr. Unfortunately, there are no
experimental data on the local moment on Cr. This is in contrast to
the much better reproduction by the TB-LMTO-ASR of the local moment on Ni in Ni-based alloys,
as compared to the CPA. This is because the fragile moment of Ni is dependent
strongly on its immediate neighbourhood, which cannot be adequately described
by the single-site CPA \cite{ni}. The moment on Cr is less dependent on
the configuration of its immediate neighbourhood and the CPA
here is not a bad description.

These observations can be explained with the help of inter and intra atomic charge
transfer effects. This has been shown in Fig.(\ref{charge}).
We notice that, for the entire concentration range, Fe gains electronic charge
from Cr. We observe that, for the alloy composition $x=0.3$, Fe gains $0.213$ 
electrons  as
compared to pure Fe. But the spin up band in Fe in the alloy
loses $0.126$ electrons, whereas the minority band
gains $0.34$ electrons, leading to a overall
reduction in the magnetic moment. A  closer study reveals that
among the different orbitals in the spin-up states, the $s$ and $p$ orbitals
gain charge while the $d$ orbital loses some. On the other hand for
the spin-down states all the three orbitals $s,p$ and $d$  gain charge.
For the Cr projected site the charge lost or gained  is different for
different orbitals leading to a small but finite
magnetic moment. For the spin-up states the $s$ and $d$ orbitals 
gain and the $p$ orbital  loses charge, while for spin-down states
only the $s$ orbital  gain and the $p$ and $d$ orbitals  lose charge.

For $x=0.5$ where the Cr projected magnetic moment changes its
sign, we observe that the $d$-up state has more charge than the  $d$-down  but
the $p$ and $s$-down states have more charge than their spin-up
counterparts. Due to this the overall magnetic moment on a  Cr
projected site reverses its sign. But for  higher concentrations
of Fe it is the charges in the  $d$ orbitals which decide the magnetic moment of Cr.
This interesting interplay  between inter and
intra atomic charge transfer has been captured by our analysis.
   
From Fig. \ref{dos} we see that the Fermi energy is pinned to the minimum of the 
minority spin density of states. But for $x > 0.5$ the majority  spin 
density of states is entirely filled. 
It shows that with increasing Fe concentration the additional
electrons are added mainly to the spin-up $d$ band. As we scan through the 
concentration range the Cr spin-up density of states shows major (x $>$ 0.7) variation.
This is understandable as we see that initially, for low $x$ compositions, 
Cr $d$-up  orbitals tend to acquire more charge than $d$-down ones, but
 for $x>0.5$ compositions the reverse phenomenon is observed.
In our calculation this is the critical concentration at which
Cr projected magnetic moment reverses its sign.

Figure \ref{magmom} (bottom panel) shows the variation of average magnetic moment against
the average number of electrons. This is the Slater-Pauling curve and
it shows a almost a linear variation up to $x=0.8$, above
which the Cr moment changes very rapidly to larger negative
value. This observation is in accordance with earlier
studies.

\section{Optical Conductivity}

In an earlier communication \cite{katu1} we had developed a completely self-consistent
TB-LMTO-ASR based method to study the optical
conductivity of alloys using a recursive procedure  suggested by 
Viswanathan and Muller\cite{visa1}.
Our starting point is the Kubo formula :

\begin{eqnarray*}
 \ll \chi^{\mu\nu}(t-t')\gg \phantom{XXXXXXXXXXXXXX}\\
 = \frac{i}{\hbar}\ \Theta(t-t')\ \langle\{\emptyset\}\otimes\phi\vert [\tilde{\j}^\mu(t) \tilde{\j}^\nu(t')]\vert\phi\otimes \vert\{\emptyset\}\rangle  
\end{eqnarray*}

\n $\tilde{\bf j}^\mu$ is the current operator and $\Theta$ is the Heaviside step function.
For disordered systems, the Hamiltonian and current operators are constructed in the full
configuration space and the Augmented Space theorem tells us that a specific matrix 
element in this space is the configuration average \cite{am1}. 
If the underlying lattice has cubic symmetry, $\chi^{\mu\nu}$ = $\chi\ \delta_{\mu\nu}$.
The fluctuation dissipation theorem relates 
the imaginary part of the Laplace
transform of the generalized  susceptibility to the Laplace transform of a correlation
function given by

\begin{figure*}                                          
\centering                                                              
\resizebox{5in}{4.0in}{\includegraphics{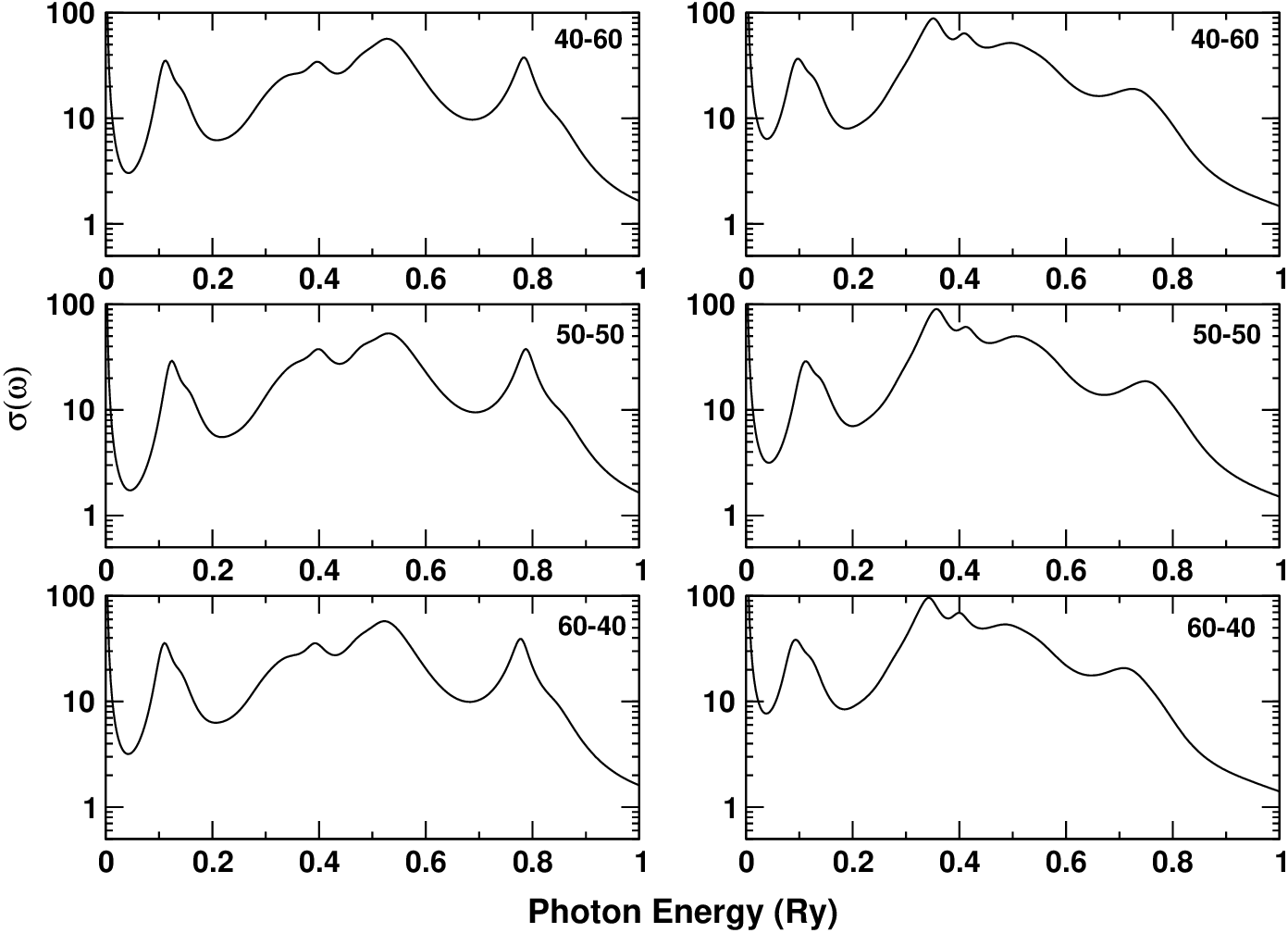}}
\vskip 5cm
\resizebox{5in}{2in}{\includegraphics{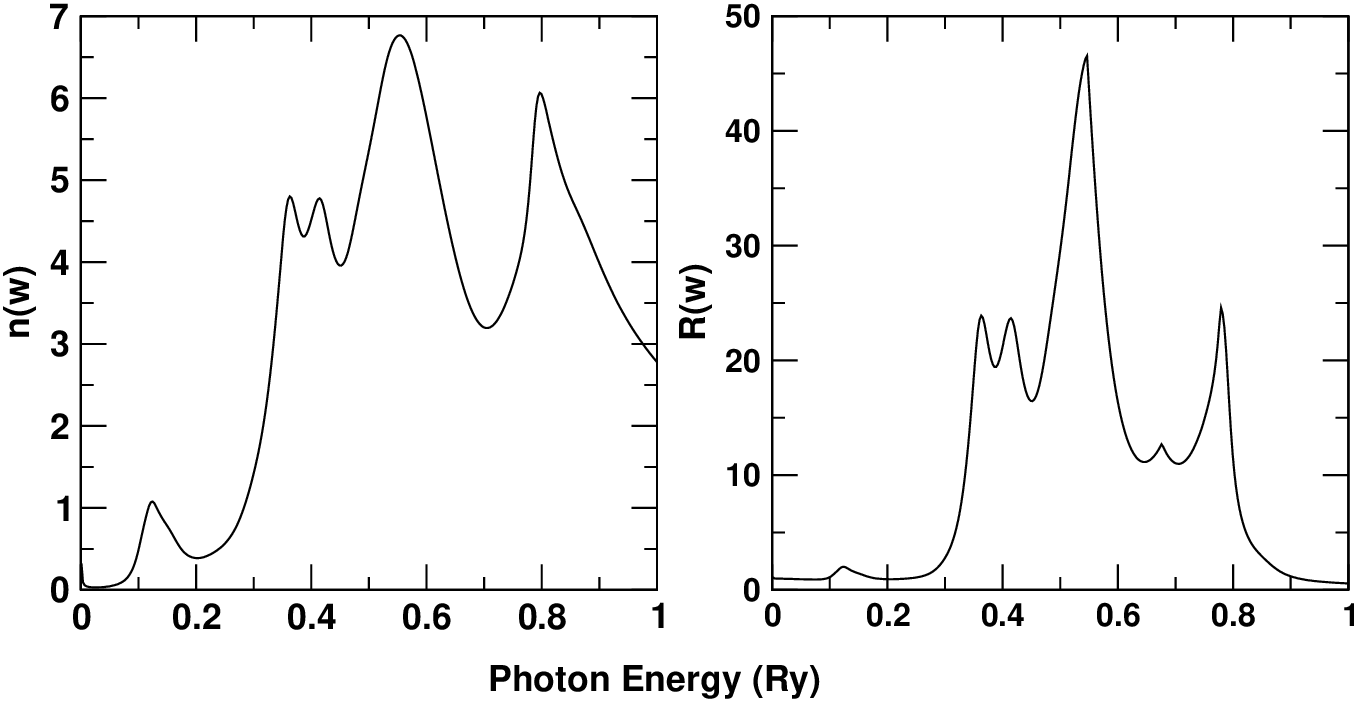}}
\caption{(top) Optical conductivity for different compositions of the FeCr alloy. Left panels are for the down-spin contributions while the right panels are
for the up-spin contributions,
(bottom) Left panel shows the refractive index $n(\omega)$ and the right panel
shows the reflectivity $R(\omega)$ for the 50-50 alloy}
\label{f.1}
\end{figure*}                                          

\begin{equation}
 \ll S(\omega)\gg  = \int_0^\infty dt  \exp\{i(\omega + i\delta)t\} \ll \mathrm{Tr}\left(\rule{0mm}{4mm} \tilde{\j}^\mu(t) \tilde{\j}^\mu(0) \right)\gg
\end{equation}

as 

\[
\ll \chi^{\prime\prime}(\omega)\gg = (1/2\hbar) \left(1-\exp\{-\beta\hbar\omega\}\right)\ll S(\omega)\gg \]
We, therefore, have to calculate the configuration average of the correlation function,
\[
\ll S(t)\gg  = \langle \{\emptyset\}\otimes\phi\vert \mathbf{j}(t)\mathbf{j}(0)\vert\phi\otimes\{\emptyset\}\rangle
\]
for a given Hamiltonian i$\widetilde{\bf H}$.
We determine the correlation directly via the recursion method suggested by Viswanath and M\"uller. 
In order to simplify the expressions for the dynamical quantities as produced by the Hamiltonian, 
we consider henceforth the modified Hamiltonian $\bar{\bf H}\ =\ \widetilde{\bf H} - E_0\tilde{\bf I}$,
 whose band energy is shifted to zero. If we start from the bra :

\[ \langle \psi(t)\vert\ =\ \langle\phi\otimes \{\emptyset\}\vert\ \tilde{\mathbf{j}}(t) \]

\n Its time evolution  is governed by the Schr\"odinger equation

\begin{equation}
 -i\ \frac{d}{dt}\left\{\rule{0mm}{4mm} \langle\psi(t)\vert\right\}\ =\ \langle\psi(t)\vert\bar{\mathbf {H}} 
\label{eq2}
\end{equation}

We now generate an orthogonal basis $\{\langle f_k\vert\}$ for representation of equation~(\ref{eq2}) in the following way . 
\begin{enumerate}
\item[(i)]We begin with initial conditions : \[ \langle f_{-1}\vert = 0\quad ;\quad \langle f_0\vert = \langle \phi\otimes \{\emptyset\}\vert{\bf j}(0)\]
\item[(ii)] We now generate the new basis members by a three term recurrence
relationship :
\[ \langle f_{k+1}\vert\ = \ \langle f_k\vert \bar{\mathbf H}\ - \ \langle f_k\vert \alpha_k \ -\ \langle f_{k-1}\vert \beta_k^2\quad\quad \mathrm {k=0,1,2}\ldots\]

\n where, \[ \alpha_k = \frac{\langle f_k \vert\bar{\bf H}\vert f_k\rangle}{\langle f_k \vert f_k\rangle}\quad\quad \beta_k^2= \frac{\langle f_k\vert f_k\rangle}{\langle f_{k-1}\vert f_{k-1}\rangle } \]
\end{enumerate}

\n We now expand the bra $\langle \psi(t)\vert $ in this orthogonal basis :

\[
 \langle \psi(t)\vert \ =\ \sum_{k=0}^\infty\ \langle f_k\vert\ D_k(t) 
\]

\n Substituting  and 
using orthogonality of the basis, we get~:

\begin{equation}
-i \dot{D}_k(t) \ =\ D_{k-1}(t)+ \alpha_k\ D_k(t) + \beta^2_{k+1}\ D_{k+1}(t)
\label{eq4}\end{equation}

\noindent with $D_{-1}(t) = 0$ and $D_k (0)=\delta_{k,0}$. We now show that the
pair of sequences generated by us, namely, $\{\alpha_k\}$ and $\{\beta_k^2\}$
are enough for us to generate the correlation function.
We note first that :

\[
 D_0(t) = \langle\psi(t)\vert f_0\rangle
 = S(t)
\]

\n Let us define the Laplace transform :

\[ d_k(z) \ =\ \int_0^\infty\ dt\ \exp{(-izt)}\ D_k(t) \]

\n Putting this back in equation(\ref{eq4}) we get :

\[
(z-\alpha_k)\ d_k(z)\ -\  i\delta_{k,0}\ =\ d_{k-1}(z)+\beta^2_{k+1}\ d_{k+1}(z)\]

k=0,1,2 $\ldots$ \\

\n This set of equations can be solved for $d_0(z)$ as a continued fraction 
representation~:

\begin{equation}
d_0(z) \ =\ \frac{i}{\displaystyle z-\alpha_0-\frac{\beta^2_1}{
\displaystyle{z-\alpha_1- \frac{\beta^2_2}{\displaystyle z-\alpha_2 - \ldots}}}}
\end{equation}

The structure function, which is the Laplace transform of the correlation function can then be obtained from the above :

\[
\ll S(\omega)\gg \ =\ \lim_{\delta\rightarrow 0}\ 2\ \Re e\ d_0(\omega+i\delta) 
\]

 The optical conductivity is then
given by :
\[ \sigma(\omega)\eq \frac{\ll S(\omega)\gg}{\omega} \]

The imaginary part of the complex dielectric function $\epsilon(\omega)$ is 
\[ \epsilon_2(\omega) \ =\ \frac{4\pi\sigma(\omega)}{\omega} \]

The real part $\epsilon_1(\omega)$ is related to the imaginary part by a 
Kramer-Kr\"onig relationship. 

The refractive index is given by :

\[ n(\omega) \ =\ \Re e \sqrt{\epsilon(\omega)} \]

Finally, the  reflectivity is defined as :

\[ R(\omega)\ =\ \left\vert \frac{\sqrt{\epsilon(\omega)}-1}{\sqrt{\epsilon(\omega)}+1}\right\vert^2 \]

Figure \ref{f.1} (top, left panel) shows the variation of the optical conductivity
for three different compositions of the FeCr alloy. For low photon energies
($\leq$ 0.05-0.06 Ry.) the behaviour is Drude like $\sim \omega^{-2}$. If we have a look at the component and spin-projected density
of states (Fig. \ref{sro1} (bottom)), we note that for these energies the transitions from below the
Fermi energy are almost entirely due to $s-p$ like states. Around 0.05-0.06 Ry 
 transitions begin from $d$-up like states coming from Cr which gives
 the characteristic peaked structures in the
density of states. Now the optical conductivity picks up and subsequent
transitions reflect the structures in the joint densities of states of the
alloy.  It is clear that this non-Drude behaviour begins for slightly larger
energies  for the down-states, so the Drude decay for the down-spin
contribution (top left panel in Fig \ref{f.1}) is more than that for
the up-spin contribution (top right panel in Fig. \ref{f.1}). 

The bottom panel of the Fig. \ref{f.1} shows the refractive index and reflectivities for the 
50-50 FeCr alloy. For low photon energies or frequencies (that is large wavelengths) the refractive index
and the reflectivities are small. However, around photon energies around 0.4, 0.6 and 0.8 Ry. the refractive
index becomes large and the reflectivity too becomes large and the system reflects back most of
the incident radiation.We can interpret this as saying that large refractive
index means that the effective speed of light in the medium becomes very small
and the photon does not propagate through the solid.  The system appears shiny for these frequency (wavelengths) and, as discussed
earlier in our work on CuZn \cite{cuzn} this determines the ``color" of the alloy.

\section{Summary}

We have used three different techniques for the calculation of the electronic structure of fully disordered \fecr alloys : the KKR-CPA, the TB-LMTO-ASR and the PAW-SQS. Each of the methods have their own
distinct approximations and the aim was to determine, for this specific alloy system, their suitability
and relative accuracy. Unlike the earlier study of Cu$_x$Zn$_{1-x}$\cite{cuzn}, we  find remarkable
agreement in the shapes of the density of states and their energy moments for all the three techniques. The local and averaged magnetic moments are also very similar, except for the Cr local  moments
in the dilute Cr limit. Unfortunately, experimental data for Cr local moment in this limit was not available. The difference in the averaged magnetic moment in the dilute limit between KKR-CPA and TB-LMTO-ASR is too small to warrant comment. Neutron scattering experiments on \fecr alloys indicate a degree
of short-ranged ordering. The single site CPA cannot effectively deal with short-ranged order. We have studied the effect of short-ranged order on the density of states of Fe$_{0.5}$Cr$_{0.5}$
via a generalization of the TB-LMTO-ASR. Finally we have extended
the TB-LMTO-ASR to study optical conductivity, reflectivity and refractive index for the same alloy system. These studies, coupled with our earlier study of the Cu$_x$Zn$_{1-x}$ alloys \cite{cuzn}, provide us with an insight into the relative merits of these approaches and allow us to choose between them for future work. 
\section*{Acknowledgments}
KT would like to thank the CSIR, India for financial assistance for research. The work was done under the Asia-Sweden Research Links 
Program titled {\sl Theoretical and experimental investigations on magnetic alloys}. BS and OE acknowledge computational support from Swedish
 National Infrastructure for Computing (SNIC). AM would like to thank the Theoretical Magnetism Group of the Department of Physics, Uppsala
University for hospitality during the time this work was completed. Part of the work was done with the VASP code and the Stuttgart TB-LMTO code was used as the basis of the ASR work.


\begin{thebibliography}{99}
\bibitem{her} A complex function $f(z)$ is called {\sl Herglotz} if (a) its
singularities lie on the real $z$ axis (b) $Sgn(Im f(z))=-Sgn(Im z)$ and (c)
$f(z)\sim 1/z$ as $z\rightarrow \pm\infty+i0$.
\end{thebibliography}
\end{document}